\documentclass[twoside,leqno,twocolumn]{article}

\usepackage{ltexpprt}

\newcommand{\Lovasz}{Lov\'asz{}}
  
\newcommand\R{{\rm I\kern-0.2em\rm R}}
\newcommand\itR{{\it I\kern-0.35em\it R}}
\newcommand\Z{{\rm Z\kern-0.2em\rm Z}}
\newcommand\itZ{{\it Z\kern-0.35em\it Z}}

\newcommand{\qed}{\hfill \mbox{$\diamond$}}

\newcommand{\term}{}

\newcommand{\mathfn}[1]{\mathop{\rm #1}}
\newcommand{\maximize}{\mathfn{maximize}}
\newcommand{\minimize}{\mathfn{minimize}}
\newcommand{\E}{\mathfn{E}}

\newcommand{\floor}[1]{\left\lfloor#1\right\rfloor}
\newcommand{\ceil}[1]{\left\lceil#1\right\rceil}

\newcommand{\bigceil}[1]{\big\lceil#1\big\rceil}

\newenvironment{tabAlgorithm}[1]{
  \setcounter{algorithmLine}{1}
  \samepage
  \begin{tabbing}
    99. \=\hspace{3ex}\=\hspace{3ex}\=\hspace{3ex}\=\hspace{3ex}\=\hspace{3ex}\kill
    #1  
    }{
  \end{tabbing}
  }
\newcounter{algorithmLine}
\newcommand{\algline}{\\\thealgorithmLine. \>\stepcounter{algorithmLine}}
\newcommand{\algnono}{\\\>}

\begin{document}

\title{
  \Large Randomized Rounding without Solving the Linear Program
\small{}\\~
\\{\it
  ((c) Sixth ACM-SIAM Symposium on Discrete Algorithms (SODA95))}
}
\author{Neal E. Young\thanks{AT\&T Bell Labs,
    rm.~2D-145, 600 Mountain Ave., Murray Hill, NJ 07974.
    Part of this research was done while at
    School of ORIE, Cornell University, Ithaca NY 14853
    and supported by \'Eva Tardos' NSF PYI grant DDM-9157199.
    E-mail: {\tt ney@research.att.com}.}
  }

\date{}
\maketitle

\pagestyle{myheadings}
\markboth{Young}{Randomized Rounding without Solving the Linear Program}
                                        

\begin{abstract} \small\baselineskip=10pt
  We introduce a new technique called oblivious rounding ---
  a variant of randomized rounding
  that avoids the bottleneck of first solving the linear program.
  Avoiding this bottleneck yields more efficient algorithms
  and brings probabilistic methods to bear on a new class of problems.
  We give oblivious rounding algorithms that approximately solve
  general packing and covering problems,
  including a parallel algorithm to find sparse strategies
  for matrix games.
\end{abstract}

\section{Introduction }

\paragraph{Randomized Rounding: }
Randomized rounding \cite{RaghavanT87}
is a probabilistic method \cite{Spencer87,AlonSE92}
for the design of approximation algorithms.
Typically, one formulates an NP-hard problem as an integer linear program,
disregards the integrality constraints,
solves the resulting linear program,
and randomly rounds each coordinate of the solution up or down
with probability depending on the fractional part.
One shows that, with non-zero probability,
the rounded solution approximates the optimal solution.
This yields a randomized algorithm;
in most cases it can be derandomized
by the method of conditional probabilities \cite{Raghavan88}.
The probabilistic analyses are often simple,
relying on just a few basic techniques.
Yet for many NP-hard problems,
randomized rounding yields the best approximation known
by any polynomial time algorithm
\cite{BertsimasV94}.

\paragraph{Oblivious Rounding: }
Derandomized or not, a main drawback of randomized rounding algorithms
has been that they first solve a linear program to find a solution to round.
We show that this bottleneck can sometimes be avoided as follows:
(1) show that randomly rounding an optimal solution
(possibly to smaller-than-integer units)
yields an approximate solution;
(2) apply the method of conditional probabilities,
finding pessimistic estimators \cite{Raghavan88}
that are essentially {\em independent}\/ of the optimal solution.
The method of conditional probabilities is used
not to derandomize per se, but to achieve the independence.

\paragraph{Generalized Packing and Covering: }
The resulting algorithms find the approximate solution
without first computing the optimal solution.
This allows randomized rounding to give simpler and more efficient algorithms
and makes it applicable for integer {\em and}\/ non-integer linear programming.
To demonstrate this, we give approximation algorithms
for general packing and covering problems
corresponding to integer and non-integer linear programs of small {\em width},
including a parallel algorithm for finding sparse, near-optimal strategies
for zero-sum games.

Packing and covering problems have been extensively studied
(see \S\ref{related-work}).
For example, Plotkin, Shmoys, and Tardos \cite{PlotkinST91}
approached these problems using Lagrangian-relaxation techniques directly.
Their algorithms and ours share the following features:
(1) they depend similarly on the width,
(2) they are Lagrangian-relaxation algorithms,
(3) they allow the packing or covering set
to be given by a (possibly approximate) subroutine for optimizing over it,
(4) they produce dual solutions that prove near-optimality,
and (5) they can provide integer solutions comparable
to those obtainable by randomized rounding.
Our approach shows a strong connection
between probabilistic techniques and Lagrangian relaxation.
Our algorithms are also relatively simple,
although they are not as effective for some problems of large width.

\paragraph{Flavor of Oblivious Rounding Algorithms: }
For the (integer) set cover problem, oblivious rounding yields
the greedy set cover algorithm \cite{Johnson74,Lovasz75}.
For the {\em fractional}\/ set cover problem, it yields an algorithm
that repeatedly chooses a set whose elements have the largest net weight,
where the weight of an element is initially $1$
and is multiplied by $1-\epsilon$ each time a set containing it is chosen.
To obtain the final cover, each set is assigned a weight proportional
to the number of times it was chosen
(this is similar in spirit to \cite{BronnimanG94} and related works).
For multicommodity flow,
it yields algorithms that repeatedly augment flow along a shortest path,
where the length of an edge is initially $1$
and is multiplied by $1+\epsilon c/c(e)$ each time the edge is used
($c(e)$ is the capacity of the edge
and $c$ is the minimum edge capacity).

\newpage
\paragraph{Problem Definitions: }
Let $P$ be a convex set in $\R^n$
and let $f$ be a linear function (not nec.~homogenous) from $P$ to $\R^m$.
The {\em width}\/ of $P$ with respect to $f$
is $\omega = \max_{j,x} f_j(x) - L$, where $L=\min_{j,x} f_j(x)$.

The {\em generalized packing problem}\/
is to compute $\lambda^* = \min_{x\in P} \max_j f_j(x)$.
The {\em packing problem}\/ occurs when $f$ is non-negative on $P$.
The {\em covering problem}\/ is to compute
$\lambda^* = \max_{x\in P} \min_j f_j(x)$, assuming $f$ is non-negative.
(This is equivalent to the generalized packing problem
with the restriction that $f$ is non-positive.)

Our algorithms assume an {\em optimization oracle}\/ for $P$ and $f$ ---
given non-negative $y\in R_m$,
the oracle returns $x$ and $f(x)$, where $x$ minimizes $\sum_j y_j f_j(x)$.
(This models, e.g., packing source-sink paths subject to edge constraints;
in this case the oracle would compute a shortest path
for given non-negative edge lengths.)
For covering, the oracle must {\em maximize}\/ the sum.

\paragraph{Quality of Solutions: }
Given the oracle, $n$, $m$, $\omega$, $L$, and $\epsilon > 0$,
our algorithms return $\epsilon$-approximate solutions.
For generalized packing, $\epsilon$
is the additive error with respect to $\lambda^*$.
For packing and covering, the error is a factor of $1\pm\epsilon$.

\paragraph{Complexity: }
Table~\ref{results} shows the number of iterations required
and the complexity per iteration.  In that caption, ``explicitly given''
means that $f(x) = Ax + b$, where $A$ and $b$ are, respectively, 
an explicitly given matrix and vector,
while $P=\{x\in \R^n : x \ge 0; \sum x_i = 1\}$.

\paragraph{Granularity: }
The oracle is called once in each iteration of the algorithm;
the algorithm returns the average of the solutions returned by the oracle.
Thus, the granularity of the final solution
is the granularity of the solutions returned by the oracle,
divided by the number of iterations.
For the abstract problems we consider,
this can provide integer solutions comparable
to those obtainable by other techniques.

\paragraph{Dual Solutions: }
Our algorithms maintain a dual solution, represented by a vector $y$,
initially uniform.  In each iteration, each $y_j$ is multiplied
by a factor depending on $f_j(x)$
where $x$ is the solution returned by the oracle
(e.g., for packing, $y_j$ is multiplied by $1+\epsilon f_j(x)/\omega$).
The average over all iterations of the values of these dual solutions
is $\epsilon$-optimal with respect to the value of the final (primal) solution.

\newcommand{\mathbox}[2]{%
  \raisebox{0pt}[#1][#1]{\parbox{6em}{$$\ceil{#2}$$}}}
\newcommand{\textbox}[1]{\begin{tabular}{r}#1\end{tabular}}

\begin{table}[t]
  \begin{center}
    \leavevmode
    \begin{tabular}{rl}
      generalized packing:&
      \mathbox{.2in}{\frac{\omega^2\ln(m)}{2\epsilon^2}}
      \\ packing:& \mathbox{.2in}{%
        \frac{(1+\epsilon)\omega\ln(m)}{\lambda^* b(\epsilon)}}
      \\ covering:& \mathbox{.2in}{%
        \frac{\omega\ln(m)}{\lambda^* b(-\epsilon)}}
    \end{tabular}
    \begin{tabular}{rl}
      & $b(\epsilon) := (1+\epsilon)\ln(1+\epsilon)-\epsilon$;
      \\  & $b(-\epsilon) > \frac{\epsilon^2}{2} > b(\epsilon)
      > \frac{2\epsilon^2}{4.2+\epsilon}$.
    \end{tabular}
  \end{center}
  \caption{Number of iterations.  Each iteration requires
    $O(\log m)$ time and $O(m)$ operations (on an EREW-PRAM),
    plus one oracle call.  For an explicitly given problem (no oracle),
    each iteration requires  $O(\log nm)$ time and $O(nm)$ operations.}
  \label{results}
\end{table}

\paragraph{Sparse Strategies for Zero-Sum Games: }
The explicitly given general packing problem generalizes
the problem of finding near-optimal strategies for zero-sum matrix games:
$P$ is the set of mixed strategies for one player,
$f_j(x)$ is the expected payoff if the player plays according to $x$
and the opponent plays the pure strategy $j$,
and $\lambda^*$ is the value of the game.
An approximate solution is a mixed strategy $x$ guaranteeing an expected payoff
within an additive $\epsilon$ of optimal.

\newpage
Each iteration chooses the best pure strategy
given that the opponent plays the mixed strategy represented by $y$.
The final solution returned is a mixed strategy that plays uniformly
 from $\bigceil{\frac{\omega^2\ln m}{2\epsilon^2}}$ pure strategies,
one for each iteration.
(The opponent has $m$ pure strategies;
$\omega$ is the maximum minus the minimum payoff.)
The existence of such sparse, near-optimal strategies 
was shown probabilistically \cite{Althofer94,LiptonY94};
our existence proof of the approximate solution
for generalized packing is a generalization of the proof in \cite{LiptonY94}.

\section{Related Work} \label{related-work}
Plotkin, Shmoys, and Tardos \cite{PlotkinST91}
(generalizing a series of works on multicommodity flow
\cite{ShahrokhiM90,KleinPST94,LeightonMPSTT91})
gave approximation algorithms for general packing and covering problems
similar to those we consider.  For these abstract problems, their results
are comparable to those in this paper, but for many problems their
results are stronger.  Most importantly, they give techniques
for reducing the effective width of a linear program
and techniques for problems (such as concurrent multicommodity flow)
when the packing or covering set is a Cartesian product.

Luby and Nisan \cite{LubyN93} give a parallel approximation algorithm
for {\term positive linear programming} ---
the special cases of linear programming
of the form $\max_x \{c\cdot{}x : Ax \le b; x \ge 0\}$ (a packing problem),
or the dual $\min_y \{b\cdot{}y : A^Ty \ge c; y \ge 0\}$ (a covering problem),
where $A$, $b$, and $c$ have non-negative coefficients.
Here $A$, $b$, and $c$ are explicitly given.

Previous algorithms applicable to zero-sum games
either required the solution of a linear program
\cite{HofmeisterL94} or did not provide sparse strategies
\cite{GrigoriadisK92,GrigoriadisK94,LubyN93}.

\newpage
\section{Introductory Example: Set Cover}
To introduce oblivious rounding, we give a simple example.
The {\term set cover problem} is the following:
given a family of sets
${\cal F}=\{S_1,\ldots,S_m\}$, with each $S_i\subseteq{\{1,2,\ldots,n\}}$,
a {\em set cover}\/ $C$ is a sub-family
such that every element $j=1,\ldots,n$ is in some set in $C$.
The problem is to find a cover $C$ that is not much larger
than $C^*$, a minimum-cardinality cover.
We derive an algorithm that, without knowing $C^*$,
emulates a random experiment that draws sets randomly from $C^*$.
The algorithm finds a cover of size at most $\ceil{|C^*|\ln n}$.

\subsection{Existence: }
Let $s=\ceil{|C^*|\ln n}$.
Consider drawing $s$ sets uniformly at random from $C^*$.
What is the expected number of elements left uncovered?
For any given element $x\in X$,
the probability that it is not covered in a given round is at most $1-1/|C^*|$,
because it is in at least one set in $C^*$.
Thus the expected number of elements left uncovered
is at most $n(1-1/|C^*|)^s < n \exp(-s/|C^*|) \le 1$.
Thus, with non-zero probability we obtain a cover $C$ of size $s$.

\subsection{Construction: }
The method of conditional probabilities naively
applied to the above proof yields an algorithm that depends on $C^*$.
We outline this next.
Our ultimate goal is not derandomization per se,
but an algorithm that does not require knowledge of $C^*$.

Consider an algorithm that chooses the sets sequentially,
making each choice deterministically
to do ``as well'' as the corresponding random choice would.
Specifically, the algorithm chooses each set
to minimize the expected number of elements
that would remain uncovered
{\em if}\/ the remaining sets were chosen randomly from $C^*$.
Letting $C$ denote the collection of sets chosen so far,
this expected number is
\begin{equation}
  \Phi(C) = 
  \sum_{j \not\in \cup C}
  \left(\frac{\sum_{S_i\not\ni j} x^*_i}{|C^*|}\right)^{s-|C|}
  \label{eqn:set-cover-pot}
\end{equation}
(We use $x^*_i$ to denote $1$ if $S_i\in C^*$ and $0$ otherwise;
we use $\cup C$ to denote the union of sets in $C$.)
$\Phi$ is called a {\em pessimistic estimator} \cite{Raghavan88},
because (a) it is an upper bound on the conditional probability of failure
(in this case, by Markov's inequality),
(b) it is initially less than $1$, and
(c) each choice can be made without increasing it.
(The latter property follows in this case
because $\Phi$ is an expected value conditioned on the choices made so far.)
These three properties imply the invariant
that if the remaining $s-|C|$ sets were to be chosen randomly from $C^*$,
the probability of failure would be less than one.
Consequently, when $|C|=s$, $C$ is a cover.

\newpage
\paragraph{Achieving Obliviousness: }
Because an uncovered element that occurs in several sets in $C^*$
contributes less to $\Phi$,
the above algorithm depends on
the number of times each element is covered by $C^*$.
This is counter-intuitive, in that the only aspect of $C^*$ used in the proof
was $\sum_{S_i \not\ni j} x^*_i/|C^*| \le 1-1/|C^*|$.
Replacing each corresponding term in $\Phi$ yields
\begin{equation}
  \tilde\Phi(C) = 
  \sum_{j \not\in \cup C}
  \left(1-\frac{1}{|C^*|}\right)^{s-|C|}.
  \label{eqn:set-cover-est}
\end{equation}
$\tilde\Phi$ is a pessimistic estimator.
More importantly, among collections of sets of the same size,
$\tilde\Phi$ is uniformly proportional
to the number of uncovered elements in the set.
Thus, the algorithm that uses $\tilde\Phi$ instead of $\Phi$
does not depend on $C^*$, it simply chooses each set
to minimize the number of elements remaining uncovered.
Nonetheless, it is guaranteed
to keep up with the random experiment,
finding a cover within $\ceil{|C^*|\ln n}$ steps.
This is the greedy set cover algorithm,
originally analyzed non-probabilistically
by Johnson~\cite{Johnson74} and \Lovasz~\cite{Lovasz75}.

\paragraph{Versus fractional cover: }
If the cover $C^*$ is a fractional cover,
the analyses of both algorithms carry over directly
to show a $\ln n$ performance guarantee.

\paragraph{What enables oblivious rounding? }
We call such algorithms {\em oblivious rounding}\/ algorithms.
What kinds of randomized rounding schemes admit them?
The key property is that the proof bounds the probability
of failure by the expected value of a sum of products
and bounds the terms corresponding across products uniformly.
To illustrate, here is the explicit proof that
\begin{math}
  \min_i \tilde\Phi(C \cup \{S_i\}) \le \tilde\Phi(C):
\end{math}
\begin{eqnarray*}
  \tilde\Phi(C) & = & \sum_{j\not\in\cup C}{
    \left(1-\frac{1}{|C^*|}\right)^{s-|C|}}
  \\ & \ge & \sum_{j\not\in\cup C}{
    \bigg(\sum_{S_i\not\ni j} \frac{x_i^*}{|C^*|}\bigg)
      \cdot \left(1-\frac{1}{|C^*|}\right)^{s-|C|-1}}
  \\ & = & \sum_i \frac{x_i^*}{|C^*|} \sum_{j\not\in(\cup C) \cup S_i}{
    \left(1-\frac{1}{|C^*|}\right)^{s-|C|-1}}
  \\ & = & \sum_i \frac{x_i^*}{|C^*|} \tilde\Phi(C \cup\{S_i\})
  \\ & \ge & \min_i \tilde\Phi(C \cup\{S_i\}).
\end{eqnarray*}
The first inequality is obtained by ``undoing''
one step of the substitution that yielded $\tilde\Phi$ from $\Phi$.
The standard argument then applies.
We use this principle for each of our analyses.

\newpage
\section{Algorithm for Generalized Packing} \label{gen-packing-sec}
Fix an instance $(P,f,L,\omega,\epsilon)$ of the generalized packing problem.
We consider randomly rounding an optimal solution
to obtain an $\epsilon$-approximate solution;
we then derive the algorithm that finds such a solution.

\subsection{Existence: }
Let $\lambda^*$ and $x^*$ be an optimal solution.
Let $S$ be a multiset obtained by repeatedly choosing random elements of $P$,
where each random element is chosen from a distribution over $P$
with $n$-dimensional mean $x^*$.
Let $\bar{x}$ be the average of the points in $S$.
\begin{lemma}  \label{gen-packing-existence}
  The probability that $\bar{x}$ is not an $\epsilon$-approximate solution
  is less than $m/\exp\big[\frac{2|S|\epsilon^2}{\omega^2}\big]$.
\end{lemma}
\begin{proof}
  Without loss of generality, assume $L=0$ and $\omega=1$.
  Otherwise take $f(x)\leftarrow\frac{f(x)-L}{\omega}$
  and $\epsilon\leftarrow\epsilon/\omega$.
  
  The convexity of $P$ ensures that $\bar{x}\in P$.
  For each $j$, $f_j(\bar{x}) = \sum_{x\in S} f_j(x)/|S|$,
  which is the average of $|S|$ independent random variables in $[0,1]$.
  Since $\E[f_j(x)]=f_j(x^*)\le \lambda^*$,
  by Hoeffding's bound \cite{Hoeffding63},
  $\Pr[f_j(\bar{x}) \ge \lambda^*+\epsilon]$
  is less than $1/\exp(2|S|\epsilon^2)$.
  Since $j$ ranges from $1$ to $m$, the result follows.
  \qed
\end{proof}

\subsection{Construction: } \label{gen-packing-const-sec}
As in the set cover example, our algorithm mimics the random experiment.
Each round it adds an element to $S$ to minimize a pessimistic estimator.
This pessimistic estimator is implicit in the existence proof.
To find it, we need the inequalities that prove (a simplified version of)
Hoeffding's bound:
\begin{lemma}[\cite{Hoeffding63}]
  Let $X=\sum X_i$ be the sum of $s$ independent random variables
  in $[0,1]$, with $\E(X_i) \le \mu_i$ and $\sum \mu_i = \mu$.
  Then $\Pr[X \ge \mu+s\epsilon] < 1/\exp(2s\epsilon^2)$.
  \label{hoeffding-lemma}
\end{lemma}
\begin{proof}
  Let $\alpha = e^{4\epsilon}-1$. 
  \begin{eqnarray}
    \lefteqn{\Pr\Big[\sum X_i \ge \mu + s\epsilon\Big]}
    \nonumber
    \\ & = & \Pr\bigg[\prod_i \frac{(1+\alpha)^{X_i}
      }{(1+\alpha)^{\mu_i+\epsilon}} \ge 1 \bigg]
    \nonumber
    \\ & \le & \E\bigg[\prod_i \frac{1+\alpha X_i
      }{(1+\alpha)^{\mu_i+\epsilon}}\bigg]
    \label{hoeffding-est}
    \\ & = & \prod_i \frac{1+\alpha E(X_i)
      }{(1+\alpha)^{\mu_i+\epsilon}}
    \nonumber
    \\ & \le & \prod_i \frac{1+\alpha \mu_i
      }{(1+\alpha)^{\mu_i+\epsilon}}
    \nonumber
    \\ & < & \prod_i 1/e^{2\epsilon^2}.
    \nonumber
  \end{eqnarray}
  The second step follows from $(1+\alpha)^z \le 1+\alpha z$
  for $0\le z \le 1$ and Markov's inequality.
  The last step uses $1+\alpha z < (1+\alpha)^{z+\epsilon}/e^{2\epsilon^2}$
  for $\epsilon > 0$, $\alpha = e^{4\epsilon}-1$, and $z \ge 0$. \qed
\end{proof}
The proof of Lemma (\ref{gen-packing-existence})
bounds the probability of failure by a sum of probabilities,
each of which is bounded
by an expected value (\ref{hoeffding-est}) in Hoeffding's proof.
Thus (when $L=0$ and $\omega=1$),
the proof bounds the probability of failure by the expected value of
\begin{displaymath}
  \sum_j \prod_{x\in S} \frac{1+\alpha f_j(x)
  }{(1+\alpha)^{\lambda^*+\epsilon}},
\end{displaymath}
the expectation of which is less than $m/\exp(2|S|\epsilon^2)$.
The conditional expectation of the sum given $T \subseteq S$ is
\begin{displaymath}
  \sum_j \bigg[\prod_{x\in T} \frac{1+\alpha f_j(x)
    }{(1+\alpha)^{\lambda^*+\epsilon}}\bigg]
  \cdot \bigg[\frac{1+\alpha f_j(x^*)]
    }{(1+\alpha)^{\lambda^*+\epsilon}}\bigg]^{s-|T|}
\end{displaymath}
where $s$ is the desired size of $S$.
To obtain the pessimistic estimator for the algorithm,
replace each $f_j(x^*)$ by the upper bound $\lambda^*$:
\begin{displaymath}
  \sum_j \bigg[\prod_{x\in T} \frac{1+\alpha f_j(x)
    }{(1+\alpha)^{\lambda^*+\epsilon}}\bigg]
  \cdot \bigg[\frac{1+\alpha \lambda^*
    }{(1+\alpha)^{\lambda^*+\epsilon}}\bigg]^{s-|T|}
\end{displaymath}
When $s$ is large enough that $m/\exp(2|S|\epsilon^2) \le 1$,
this quantity is a pessimistic estimator:
(a) it is an upper bound on the conditional probability of failure,
(b) it is initially less than $1$,
and (c) some $x$ can always be added to $S$ without increasing it. 
Properties (a) and (b) follow from the choice of $s$
and the inequalities in the proof of Hoeffding's lemma.
Property (c) follows from the derivation,
as explained for the set cover example.
Among multisets of a given size,
this pessimistic estimator is uniformly proportional to
\begin{displaymath}
  \sum_j \prod_{x\in T} 1+\alpha f_j(x).
\end{displaymath}
Thus, to augment a given multiset $T$, the algorithm adds the element $x$
minimizing $\sum_j y_j f_j(x)$,
where $y_j = \prod_{x\in T} 1+\alpha f_j(x)$.
This, accounting for the normalization $L=0$ and $\omega=1$,
is the algorithm in Figure~\ref{gen-packing-fig}.

\begin{figure}[t]
  \begin{tabAlgorithm}{%
      $\mbox{\sc Find-Generalized-Packing}(P,f,L,\omega,\epsilon)$}
    \algline $\epsilon \leftarrow \frac{\epsilon}{\omega}$;
    $\alpha \leftarrow e^{4\epsilon}-1$;
    $S \leftarrow \{\}$;
    $s \leftarrow \frac{\ln m}{2\epsilon^2}$
    \algline $y_j \leftarrow 1$~~($j=1,\ldots,m$)
    \algline {\bf repeat}
    \algline    \> choose $x\in P$ to minimize $\sum_j y_j f_j(x)$
    \algline    \> $S\leftarrow S\cup \{x\}$
    \algline    \> $y_j \leftarrow y_j\cdot{
      \big[1+\alpha\frac{f_j(x)-L}{\omega}\big]}$~~($j=1,\ldots,m$)
    \algline {\bf until} $|S| \ge s$
    \algline {\bf return} $\frac{1}{|S|}\sum_{x\in S} x$
  \end{tabAlgorithm}
  \caption{Algorithm for generalized packing}
  \label{gen-packing-fig}
\end{figure}

\section{Packing and Covering Algorithms} \label{packing-sec}
We derive the packing algorithm analogously.
Fix an instance $(P,f,\omega,\epsilon)$ of the packing problem.
Let $\lambda^*$, $x^*$, $S$ and $\bar{x}$
be as for Lemma~\ref{gen-packing-existence}.
Note that, for this problem, an $\epsilon$-approximate solution
is an $x\in P$ with $f(x) \le (1+\epsilon)\lambda^*$.

\subsection{Existence: }
\begin{lemma}  \label{packing-existence}
  The probability that $\bar{x}$ is not an $\epsilon$-approximate
  solution is less than
  $m/\exp\left[\frac{|S|b(\epsilon)\lambda^*}{\omega}\right]$.
\end{lemma}
\begin{proof}
  Without loss of generality, assume $\omega=1$.  Otherwise take
  $f(x)\leftarrow f(x)/\omega$ and $\lambda^*\leftarrow\lambda^*/\omega$.
  
  The convexity of $P$ ensures that $\bar{x}\in P$.
  For each $j$, $f_j(\bar{x}) = \sum_{x\in S} f_j(x)/|S|$,
  which is the average of $|S|$ independent random variables in $[0,1]$,
  each with expectation $f_j(x^*) \le \lambda^*$.
  By Raghavan's bound \cite{Raghavan88},
  $\Pr[f_j(\bar{x}) \ge (1+\epsilon)\lambda^*]$
  is less than $1/\exp[|S|b(\epsilon)\lambda^*]$.
  Since $j$ ranges from $1$ to $m$, the result follows.
  \qed
\end{proof}

\subsection{Construction: }
Here is Raghavan's proof:
\begin{lemma}[\cite{Raghavan88}]
  Let $X=\sum X_i$ be the sum of independent random variables in $[0,1]$
  with $\E(X_i) \le \mu_i$ and $\sum \mu_i = \mu > 0$.

  Then $\Pr[X \ge (1+\epsilon)\mu] < 1/\exp[b(\epsilon)\mu]$.
  \label{raghavan-lemma}
\end{lemma}
\begin{proof}
  \begin{eqnarray}
    \lefteqn{\Pr[X \ge (1+\epsilon)\mu]}
    \nonumber
    \\ & = & \Pr\bigg[\prod_i \frac{(1+\epsilon)^{X_i}
      }{(1+\epsilon)^{(1+\epsilon)\mu_i}} \ge 1 \bigg]
    \nonumber
    \\ & \le & \E\bigg[\prod_i \frac{1+\epsilon X_i
      }{(1+\epsilon)^{(1+\epsilon)\mu_i}}\bigg]
    \label{raghavan-est}
    \\ & = &  \prod_i \frac{1+\epsilon \E(X_i)
      }{(1+\epsilon)^{(1+\epsilon)\mu_i}}
    \nonumber
    \\ & < & \prod_i \frac{e^{\epsilon \mu_i}
      }{(1+\epsilon)^{(1+\epsilon)\mu_i}}
    \nonumber
  \end{eqnarray}
  The last line equals $1/\exp[b(\epsilon)\mu]$.
  The second step uses $(1+\alpha)^z \le 1+\alpha z$
  for $0\le z \le 1$ and Markov's inequality.
  The last uses $\E(X_i)\le \mu_i$
  and  $1+z \le e^z$, which is strict if $z\ne 0$.
  \qed
\end{proof}
Thus (assuming $\omega=1$), the proof of Lemma \ref{packing-existence}
bounds the probability of failure by the expectation of
\begin{displaymath}
  \sum_j \prod_{x\in S}
  \frac{1+\epsilon f_j(x)
    }{(1+\epsilon)^{(1+\epsilon)\lambda^*}}
\end{displaymath}
corresponding to (\ref{raghavan-est}).
The expectation given $T\subseteq S$ is
\begin{displaymath}
  \sum_j \bigg[\prod_{x\in T}
  \frac{1+\epsilon f_j(x)
    }{(1+\epsilon)^{(1+\epsilon)\lambda^*}}\bigg]
  \cdot\bigg[\frac{1+\epsilon f_j(x^*)
    }{(1+\epsilon)^{(1+\epsilon)\lambda^*}}\bigg]^{s-|T|},
\end{displaymath}
where $s$ is the desired size of $S$.
When $s$ is large enough that
$m/\exp[|S|b(\epsilon)\lambda^*] \le 1$,
replacing $f_j(x^*)$ by $\lambda^*$
gives a pessimistic estimator.
Among multisets $T$ of the same size,
the pessimistic estimator is proportional to
\begin{displaymath}
  \sum_j \prod_{x\in T} 1+\epsilon f_j(x).
\end{displaymath}
Thus, to augment a given multiset $T$,
the algorithm adds the element $x$ minimizing
\begin{math}
  \sum_j y_j f_j(x),
\end{math}
where $y_j = \prod_{x\in T} 1+\epsilon f_j(x)$.
This, accounting for the normalization to the case $\omega=1$,
gives the algorithm in Figure~\ref{packing-fig-s}.
This algorithm assumes $s$ is given.
We remove this requirement in Section~\ref{dual-sec}.

\begin{figure}[t]
  \begin{tabAlgorithm}{$\mbox{\sc Find-Packing-Given-$s$
        }(P,f,\epsilon,\omega,s)$}
    \algline $S \leftarrow \{\}$
    \algline $y_j \leftarrow 1$~~($j=1,\ldots,m$)
    \algline {\bf repeat}
    \algline    \> choose $x\in P$ to minimize $\sum_j y_j f_j(x)$
    \algline    \> $S\leftarrow S\cup \{x\}$
    \algline    \> $y_j \leftarrow y_j\cdot{
      \big[1+\epsilon \frac{f_j(x)}{\omega}\big]}$~~($j=1,\ldots,m$)
    \algline {\bf until} $|S|\ge s$
    \algline {\bf return} $\frac{1}{|S|}\sum_{x\in S} x$
  \end{tabAlgorithm}
  \caption{Algorithm for packing, given $s$. To obtain covering algorithm,
    negate $\epsilon$ and change ``$\minimize$'' to ``$\maximize$''.}
  \label{packing-fig-s}
\end{figure}

\subsection{Covering Algorithm. }
The covering algorithm is described in Figure~\ref{packing-fig-s}.
Its derivation is analogous to that of the packing algorithm.
Fix an instance $(P,f,\omega,\epsilon)$ of the approximate covering problem.
Let $\lambda^*$, $x^*$, $S$ and $\bar{x}$
be as for Lemma~\ref{gen-packing-existence}.
Note that, for this problem, $\lambda^* = \min_j f_j(x^*)$
and an $\epsilon$-approximate solution $x\in P$
satisfies $f(x) \ge (1-\epsilon)\lambda^*$.
\begin{lemma}
  The probability that $\bar{x}$ is not an $\epsilon$-approximate solution
  is less than $m/\exp[|S|b(\epsilon)\lambda^*/\omega]$. \hfill
  \label{covering-existence}
\end{lemma}
We omit the proof, which is essentially the same as for packing,
except it is based on the following variant of Raghavan's bound:
\begin{lemma}[\cite{Raghavan88}] \label{raghavan-variant}
  Let $X=\sum X_i$ be the sum of independent random variables in $[0,1]$
  with $\E(X_i) \ge \mu_i$ and $\sum \mu_i = \mu > 0$.

  Then $\Pr[X \le (1-\epsilon)\mu] < 1/\exp[b(-\epsilon)\mu]$.
\end{lemma}
We omit the derivation of the algorithm,
noting only that the proof of Lemma~\ref{covering-existence}
implicitly bounds the probability of failure by the expectation of
\begin{displaymath}
  \sum_j \prod_{x\in S}\frac{1-\epsilon f_j(x)
    }{(1-\epsilon)^{(1-\epsilon)\lambda^*}}.
\end{displaymath}

\section{Dual Solutions} \label{dual-sec}
Our algorithms implicitly find good approximate solutions
to the underlying dual linear programs.
The argument that the algorithm ``keeps up''
with the random rounding of an unknown optimal solution
implicitly uses a dual solution to bound the optimal at each iteration.
The value of the solution generated by the algorithm
thus converges not only to the value of the optimum,
but also to the average of the values of these dual solutions.
The basic principle in each case is similar to that for set cover,
which we give first for illustration.

\subsection{Set Cover Dual: }
The dual problem is to assign non-negative weights to the elements
so that the net weight assigned to the elements in any set is at most one.
The value of the dual solution is the net weight assigned.

At the start of a given iteration,
suppose $r$ elements remain uncovered,
and let $d$ denote the largest number in any set in $\cal F$.
Then assigning each uncovered element a weight of $1/d$
yields a dual solution of value $v=r/d$.

During the course of the algorithm,
let $\bar{v}$ denote the harmonic mean of the dual solutions
corresponding to the iterations so far.

\begin{lemma}
  The set cover algorithm maintains the invariant that
  the number of elements not covered by the current partial cover $C$
  is less than $n/\exp(|C|/\bar{v})$.
\end{lemma}
The proof is essentially the same as the proof
that $\tilde\Phi$ is a pessimistic estimator,
except the values of the dual solutions take the place of $|C^*|$.
\begin{proof}
  In an iteration where the dual solution has value $r/d$,
  the number of uncovered elements decreases
  from $r$ to $r-d = r(1-1/v) < r e^{-1/v}$.
  By induction on the iterations, the algorithm maintains the invariant
  that the number of uncovered elements is less than
  $n/\exp\big(\sum_\ell 1/v_\ell\big)$
  where $v_\ell$ is the value of the dual solution
  corresponding to the $\ell$th iteration
  and $\ell$ ranges over the iterations so far.
  Note that $\bar{v} = |C|/\sum \frac{1}{v_\ell}$.
  \qed
\end{proof}

Before the last iteration at least one element is left,
so at that point $n/\exp((k-1)/\bar{v}) > 1$.  Thus,
\begin{corollary}
  The harmonic mean of the values of the dual solutions
  over the first $k-1$ iterations
  is larger than \(\frac{k-1}{\ln n}\),
  where $k$ is the size of the final cover.
\end{corollary}

The maximum value is at least the arithmetic mean,
which is at least the harmonic mean,
so at least one of these simple dual solutions
has value above $\frac{k-1}{\ln n}$.

\subsection{Generalized Packing Dual: }
The vector $y$ maintained by the generalized packing
algorithm represents a dual solution.
At the start of a given iteration,
the value of the dual solution associated with $y$ is
\begin{equation}
  \frac{\min_{x\in P} \sum_j y_j f_j(x)}{\sum_j y_j}.
  \label{dual-value}
\end{equation}
(Since $y \ge 0$, a simple argument shows this is a lower bound
on $\lambda^*=\min_{x\in P}\max_j f_j(x)$.)

\paragraph{Notation: }
During the course of the algorithm,
let $\bar{x}$ denote the current solution $\sum_{x\in S} x/|S|$
represented by $S$.
Let $\bar\lambda$ denote $\max_j f_j(\bar{x})$.
Let $\bar{v}$ denote the average of the values of the dual solutions
for the previous iterations.
Let $v(x,y)$ denote $\sum_j y_j f_j(x)/\sum_j y_j$.

\begin{lemma}
  The generalized packing algorithm maintains the invariant
  \[(1+\alpha)^{|S|(\bar\lambda-L)/\omega}
  \le (1+\alpha)^{|S|((\bar{v}-L)/\omega+\epsilon)}m/\exp(2|S|\epsilon^2).\]
  \label{gen-packing-inv}
\end{lemma}
\begin{proof}
  WLOG, assume $L=0$ and $\omega=1$.
  We show that $\sum_j y_j$ is at least the left-hand side
  and at most the right-hand side.
  The first part follows from the same sequence of inequalities
  that was used in \S\ref{gen-packing-const-sec}
  to derive the (numerator of the) pessimistic estimator:
  \begin{eqnarray*}
    (1+\alpha)^{|S|\bar\lambda} & \le & \sum_j (1+\alpha)^{|S|f_j(\bar{x})}
    \\ & = & \sum_j \prod_{x\in S} (1+\alpha)^{f_j(x)}
    \\ & \le & \sum_j \prod_{x\in S} 1+\alpha f_j(x).
  \end{eqnarray*}
  Since $y_j =  \prod_{x\in S} 1+\alpha f_j(x)$, the first part follows.

  For the second part,
  we first note the role of the dual solution in each iteration:
  given the current $x$ and $y$,
  the iteration increases the quantity $\sum_j y_j$
  by a factor of $1+\alpha v(x,y)$.
  (This follows from inspection of the algorithm
  and the definition of $v(x,y)$.)
  Next we apply the sequence of inequalities
  that bounded the pessimistic estimator below $1$
  in \S\ref{gen-packing-const-sec}:
  By the last inequality in Hoeffding's bound (Lemma~\ref{hoeffding-lemma}),
  $1+\alpha v(x,y) \le (1+\alpha)^{v(x,y)+\epsilon}/\exp(2\epsilon^2)$.
  Let $v_\ell$ denote the value of $v(x,y)$
  at the $\ell$th iteration (for $1 \le \ell \le |S|$).
  By induction on the iterations
  \begin{displaymath}
    \sum_j y_j
    \le m (1+\alpha)^{\sum_\ell (v_\ell+\epsilon)}/\exp(2|S|\epsilon^2).
  \end{displaymath}
  Since $|S|\bar{v} = \sum_\ell v_\ell$, this gives the result.
  \qed
\end{proof}

\begin{corollary}
  After $\bigceil{\frac{\omega^2\ln m}{2\epsilon^2}}$ iterations
  of the generalized packing algorithm,
  $\bar\lambda \le \bar{v} + \epsilon$.
  That is, the primal and average dual values differ by at most $\epsilon$.
  \label{gen-packing-corollary}
\end{corollary}

\subsection{Packing Dual: }
The packing and covering algorithms also generate implicit dual solutions
whose average values converge to the primal value.
Let $\bar{\lambda}$ and $\bar{v}$ be defined
as for the generalized packing dual.
\begin{lemma}
  The packing algorithm maintains the invariant that
  \[(1+\epsilon)^{|S|\bar\lambda/\omega}
  \le m e^{\epsilon |S|\bar{v}/\omega}.\]
  \label{packing-inv}
\end{lemma}
We omit this and subsequent proofs in this section,
since they are similar to that of Lemma~\ref{gen-packing-inv}.
\begin{corollary}
  After $\bigceil{\frac{(1+\epsilon)\omega\ln m}{\lambda^* b(\epsilon)}}$
  iterations of the packing algorithm,
  $\bar\lambda \le (1+\epsilon)\bar{v}$.
  That is, the primal and average dual values
  differ by at most a factor of $1+\epsilon$.
\end{corollary}

Our final packing algorithm detects convergence
by comparing the primal value to the best dual value so far.
The algorithm is shown in Figure~\ref{packing-fig}.
The algorithm maintains $f(\bar{x})$ (in the variable $F$)
instead of $\bar{x}$.

\begin{figure}[t]
  \begin{tabAlgorithm}{$\mbox{\sc Find-Packing}(P,f,\epsilon,\omega)$}
    \algline $S \leftarrow \{\}$; $y_j \leftarrow 1$~~($j=1,\ldots,m$)
    \algline {\bf repeat}
    \algline    \> choose $x\in P$ to minimize $v=\sum_j y_j f_j(x)$
    \algline    \> $S\leftarrow S\cup \{x\}$
    \algline    \> $y_j \leftarrow y_j\cdot{
      \big[1+\epsilon \frac{f_j(x)}{\omega}\big]}$~~($j=1,\ldots,m$)
    \algline    \> $V\leftarrow\max(V,v/\sum_j y_j)$
    \algline    \> $F_j \leftarrow [(|S|-1)F_j+f_j(x)]/|S|$~~($j=1,\ldots,m$)
    \algline    \> $\bar\lambda \leftarrow \max_j F_j$
    \algline {\bf until} $\bar\lambda \le (1+\epsilon) V$
    \algline {\bf return} $\sum_{x\in S} x/|S|$
  \end{tabAlgorithm}
  \caption{Algorithm for packing.  To obtain covering algorithm,
    negate $\epsilon$'s and change each ``$\max$'' to ``$\min$'',
    ``$\minimize$'' to ``$\maximize$'', and ``$\le$'' to ``$\ge$''.}
  \label{packing-fig}
\end{figure}

\subsection{Covering Dual: }
\begin{lemma}
  The covering algorithm maintains the invariant that
  \[(1-\epsilon)^{|S|\bar\lambda/\omega}
  \le m e^{-\epsilon |S|\bar{v}/\omega}.\]
  \label{covering-inv}
\end{lemma}

\begin{corollary}
  After $\bigceil{\frac{\omega\ln m}{\lambda^* b(-\epsilon)}}$ iterations
  of the covering algorithm, $\bar\lambda \ge (1-\epsilon)\bar{v}$, that is,
  the primal and average dual values
  differ by at most a factor of $1-\epsilon$.
\end{corollary}
The algorithm is described in Figure~\ref{packing-fig}.

\newpage
\section{Using an Approximate Oracle}
If the subroutine for computing $\min_{x\in P} \sum_j y_j f_j(x)$
returns only an {\em approximate}\/ minimizer $x$,
our algorithms still work well.
The degree of approximation (absolute and/or relative)
of the subroutine carries over into the performance guarantee of the algorithm.
For covering, it can also affect the convergence rate
(and therefore the granularity).

We model the error by assuming that, given $y$,
the oracle returns an $x$ such that
\begin{equation}
  v(x,y) \le (1+\delta_1) \min_{x\in P} v(x,y) + \delta_2
  \label{app-oracle-def}
\end{equation}
where $v(x,y) = \sum_j y_j f_j(x) / \sum_j y_j$,
$\delta_1\ge 0$ denotes the relative error
and $\delta_2\ge 0$ denotes the absolute error.
We call this a {\em ($\delta_1,\delta_2$)-approximate}\/ oracle.
(For covering, the notion of approximation is defined analogously.)

In each iteration, $y$ still represents a dual solution.
Since $x$ is only an approximate minimizer,
the value of the dual solution is no longer $v(x,y)$,
but it is at least $\frac{v(x,y)-\delta_2}{1+\delta_1}$.
Still using $\bar{v}$ to denote the average of the values
of the dual solutions for the previous iterations,
define $\tilde{v}$ to be the average of the corresponding $v(x,y)$'s.
Lemmas~\ref{gen-packing-inv},~\ref{packing-inv}, and~\ref{covering-inv}
go through directly provided ``$\tilde{v}$'' is substituted for ``$\bar{v}$''.
 From the (modified) lemmas,
by the same reasoning that gives the corollaries to those lemmas,
together with the fact that
 $\tilde{v} \le (1+\delta_1)\bar{v}+\delta_2$,
we get the following propositions.

\begin{proposition}
  Suppose the generalized packing algorithm uses
  a $(\delta_1,\delta_2)$-approximate oracle.
  After $\ceil{\frac{\omega^2\ln m}{2\epsilon^2}}$ iterations,
  $\bar\lambda
  \le \tilde{v} + \epsilon
  \le (1+\delta_1)\bar{v} + \delta_2 + \epsilon$.
  \label{gen-packing-app-corollary}
\end{proposition}

\begin{proposition}
  Suppose the packing algorithm uses 
  a $(\delta_1,\delta_2)$-approximate oracle.  After
  $\ceil{\frac{(1+\epsilon)\omega\ln m}{\lambda^* b(\epsilon)}}$
  iterations,
  $\bar\lambda
  \le (1+\epsilon)\tilde{v}
  \le (1+\epsilon)(1+\delta_1)\bar{v} + (1+\epsilon)\delta_2$.
\end{proposition}

For covering, $\tilde{v} \ge (1-\delta_1)\bar{v}-\delta_2$.
\begin{proposition}
  Suppose the covering algorithm uses 
  a $(\delta_1,\delta_2)$-approximate oracle.  After
  $$\ceil{\frac{\omega\ln m
    }{[(1-\delta_1)\lambda^*-\delta_2] b(-\epsilon)}}$$
  iterations,
  $\bar\lambda
  \ge (1-\epsilon)\tilde{v}
  \ge (1-\epsilon)(1-\delta_1)\bar{v} - (1-\epsilon)\delta_2$.
\end{proposition}

These results hold for the algorithms without modification.
In particular, $V$ in the packing algorithm in Figure~\ref{packing-fig}
equals the best $v(x,y)$ seen so far,
which is at least $\tilde{v}$,
so is guaranteed to be within a $1+\epsilon$ factor of $\bar\lambda$
within the required number of rounds.

\newpage
\section{Integer Packing and Covering}
The packing and covering algorithms in Figure~\ref{packing-fig},
as they stand, do not allow explicit control over the granularity
of the final solution.  Because the number of iterations can be
less than the upper bound, the algorithms only guarantee a lower bound
on the granularity.  Of course, the lower bound is the difficult part,
so it is not surprising that exact control over the granularity
can be obtained.
In this section, we discuss briefly how to modify those algorithms
to find, e.g., an integer solution.

For simplicity, we consider a particular case of integer packing.
Fix an instance of the packing problem $(P,f,\omega,\epsilon)$.
Let $\lambda^*$ and $x^*$ be an optimal solution.
In addition, let $V \subset P$ be the extreme points
on the boundary of $P$ (if $P$ is a polytope, $V$ is its vertex set).
We assume that the oracle returns only elements of $V$.
The {\em integer packing} problem is to compute a maximum cardinality
multiset $S \subseteq V$ such that $\sum_{x\in S} f_j(x) \le 1$.

Note that, for any such $S$, $|S| \le \floor{1/\lambda^*}$
because $f(\bar{x}) \le 1/|S|$, where $\bar{x} = \sum_{x\in S} x/|S|$.
An {\em $\epsilon$-approximate integer solution} is a set $S$
such that $|S| \ge \floor{1/\lambda^*}$
and $\sum_{x\in S} f_j(x) \le 1+\epsilon$.

Let $S$ be a multiset obtained by repeatedly choosing random elements of $V$,
where each random element is chosen from a distribution on $V$
with mean $x^*$.
(Such distributions exist because $P$ is the convex closure of $V$.)
\begin{lemma}
  When $|S| \le 1/\lambda^*$,
  $$\Pr\Big[(\exists j)~ \sum_{x\in S} f_j(x) \ge 1+\epsilon\Big]
   < m/\exp[b(\epsilon)/\omega].$$
\end{lemma}
The proof is essentially the same as that of Lemma~\ref{packing-existence},
except $1/|S|$ replaces $\lambda^*$.

A corollary to the lemma is that,
provided $m/\exp[b(\epsilon)/\omega] \le 1$,
there exists an $\epsilon$-approximate integer solution.
The corresponding algorithm is the same as the basic packing algorithm,
except the termination condition is different.
The algorithm terminates when adding another element would cause
$\sum_{x\in S} f_j(x) > 1+\epsilon$ for some $j$.
Because the algorithm keeps up with the random process,
the resulting set has size at least $\floor{1/\lambda^*}$.

\paragraph{Complexity and Performance Guarantee: }
The algorithm is given in Figure~\ref{int-packing-fig}.
Note that $\floor{1/\lambda^*} \le |S| \le \floor{(1+\epsilon)/\lambda^*}$,
so the number of iterations in this case is at most $(1+\epsilon)/\lambda^*$.
For the condition  $m/\exp[b(\epsilon)/\omega] \le 1$,
it suffices that, for instance,
$\epsilon \ge 2\max(\omega\ln m,\sqrt{\omega\ln m}).$

\paragraph{Covering: }
The same techniques apply for integer covering.
For covering, define an {\em $\epsilon$-approximate integer solution}
to be a set $S$ such that $|S| \le \ceil{1/\lambda^*}$
and $\sum_{x\in S} f_j(x) \ge 1-\epsilon$.
(Many variations are possible.)
Let $S$ be a random multiset as above.
\begin{lemma}
  When $|S| \ge 1/\lambda^*$,
  $$\Pr\Big[(\exists j)~ \sum_{x\in S} f_j(x) \le 1-\epsilon\Big]
   < m/\exp[b(-\epsilon)/\omega].$$
\end{lemma}
The resulting algorithm is described in Figure~\ref{int-packing-fig}.
The number of iterations in this case is at most $\ceil{1/\lambda^*}$.
For the condition  $m/\exp[b(-\epsilon)/\omega] \le 1$,
it suffices that $\epsilon \ge \sqrt{2\omega\ln m}.$

\begin{figure}[t]
  \begin{tabAlgorithm}{$\mbox{\sc Find-Integer-Packing
        }(P,f,\epsilon,\omega)$}
    \algnono assumption: $m/\exp[b(\epsilon)/\omega] \le 1$.
    \algline $S \leftarrow \{\}$;
    $y_j \leftarrow 1$, $F_j \leftarrow 0$~~($j=1,\ldots,m$)
    \algline {\bf repeat}
    \algline    \> choose $x\in P$ to minimize $\sum_j y_j f_j(x)$
    \algline    \> $F_j \leftarrow F_j+f_j(x)$~~($j=1,\ldots,m$)
    \algline    \> {\bf if} $\max_j F_j > 1+\epsilon$ {\bf return} $S$
    \algline    \> $S\leftarrow S\cup \{x\}$
    \algline    \> $y_j \leftarrow y_j\cdot{
      \big[1+\epsilon \frac{f_j(x)}{\omega}\big]}$~~($j=1,\ldots,m$)
  \end{tabAlgorithm}
  \caption{Algorithm for integer packing.
    To obtain covering algorithm,
    negate $\epsilon$'s and change ``$\max$'' to ``$\min$'',
    ``$\minimize$'' to ``$\maximize$'', and ``$>$'' to ``$<$''.
    }
  \label{int-packing-fig}
\end{figure}

\section{Conclusion}
\paragraph{Partial derandomization: }
The point of oblivious rounding
is not derandomization per se,
but to achieve independence
 from the unknown aspects of the optimal solution.
For some random rounding schemes,
some of the parameters of the random process are known;
these can be left in the algorithm.
For instance, in concurrent multicommodity flow,
the relative amount of flow of each commodity is known.
A natural randomized rounding scheme
is to choose a commodity with probability
proportional to its (known) demand,
and then to choose a flow path among paths for that commodity
with probability proportional to its (unknown) weight in the optimal flow.
Applying oblivious rounding to only the second random choice
gives a randomized algorithm in the style of \cite{PlotkinST91}.

\paragraph{Mixed bounds: }
Each of the random analyses in this paper
employed a single type of probabilistic bound.
This is not a limitation of the technique.
Oblivious rounding can be applied to
analyses using, e.g., sums of probabilities
bounded by Raghavan's bounds,
Hoeffding's bound, and Markov's inequality.
This is relatively straightforward,
if technically more tedious.

\paragraph{More general functions: }
Chernoff-type bounds exist for more general classes of functions
than linear functions (e.g., Azuma's inequality \cite{AlonSE92}).
A natural question is whether oblivious rounding
can be applied to such bounds to optimize more general functions.

\newcommand{\noopsort}[1]{} \newcommand{\printfirst}[2]{#1}
  \newcommand{\singleletter}[1]{#1} \newcommand{\switchargs}[2]{#2#1}

\end{document}